\documentclass[aps,superscriptaddress,prc,twocolumn,nofootinbib]{revtex4}

\usepackage{graphicx}
\usepackage{epstopdf}
\usepackage{subfigure}
\usepackage{epsfig}
\usepackage{amsmath,amssymb,amsfonts}
\usepackage{color}
\usepackage[utf8]{inputenc}

\begin{document}

\title{Heavy and light flavor jet quenching at RHIC and LHC energies}

\author{Shanshan Cao}
\affiliation{Nuclear Science Division, Lawrence Berkeley National Laboratory, Berkeley, CA 94720, USA}
\affiliation{Department of Physics and Astronomy, Wayne State University, 666 W. Hancock St., Detroit, MI 48201, USA}
\author{Tan Luo}
\affiliation{Institute of Particle Physics and Key Laboratory of Quark and Lepton Physics (MOE), Central China Normal University, Wuhan, 430079, China}
\author{Guang-You Qin}
\affiliation{Institute of Particle Physics and Key Laboratory of Quark and Lepton Physics (MOE), Central China Normal University, Wuhan, 430079, China}
\author{Xin-Nian Wang}
\affiliation{Institute of Particle Physics and Key Laboratory of Quark and Lepton Physics (MOE), Central China Normal University, Wuhan, 430079, China}
\affiliation{Nuclear Science Division, Lawrence Berkeley National Laboratory, Berkeley, CA 94720, USA}

\date{\today}

%%%%%%%%%%%%%%%%%%%%%%%%%%%%%%%%%%%%%%%%%%%%%%%%%%%%%%%%%%%%%%%%%%%%%

\begin{abstract}

The Linear Boltzmann Transport (LBT) model coupled to hydrodynamical background is extended to include transport of both light partons and heavy quarks through the quark-gluon plasma (QGP) in high-energy heavy-ion collisions. The LBT model includes both elastic and inelastic medium-interaction of both primary jet shower partons  and thermal recoil partons within perturbative QCD (pQCD). It is shown to simultaneously describe the experimental data on heavy and light flavor hadron suppression in high-energy heavy-ion collisions for different centralities at RHIC and LHC energies. More detailed investigations within the LBT model illustrate the importance of both initial parton spectra and the shapes of fragmentation functions on the difference between the nuclear modifications of light and heavy flavor hadrons. The dependence of the jet quenching parameter $\hat{q}$ on medium temperature and jet flavor is quantitatively extracted.
\end{abstract}

\maketitle

%%%%%%%%%%%%%%%%%%%%%%%%%%%%%%%%%%%%%%%%%%%%%%%%%%%%%%%%%%%%%%%%%%%%%

\textit{Introduction} ---
Jet quenching is one of the most important signatures for the formation of the hot and dense quark-gluon plasma (QGP) in heavy-ion collisions at the Relativistic Heavy-Ion Collider (RHIC) and the Large Hadron Collider (LHC) \cite{Wang:1991xy, Qin:2015srf}. One important phenomenon of jet quenching is the suppression of single inclusive hadron spectra at large transverse momentum. It is caused by energy loss of jet partons from early hard scattering via elastic (collisional) and inelastic (radiative) interactions with the color-deconfined QGP medium before fragmenting into hadrons.  Theoretical studies \cite{Gyulassy:1993hr, Baier:1996kr, Baier:1996sk, Baier:1998kq, Zakharov:1996fv, Gyulassy:1999zd, Wiedemann:2000za,Guo:2000nz,Arnold:2002ja,Armesto:2003jh, Djordjevic:2008iz, Majumder:2009ge, CaronHuot:2010bp} have shown that both elastic and inelastic parton energy loss are controlled by a series of jet transport coefficients, one of which is the jet quenching parameter $\hat{q}$, denoting the transverse momentum  transfer squared per unit time between the propagating hard partons and the soft medium. The first systematic effort to quantitatively extract $\hat{q}$ was performed by the JET Collaboration by comparing several jet quenching model calculations with the experimental data for the nuclear modification of single inclusive hadron production at high transverse momentum ($p_\mathrm{T}$) at RHIC and the LHC \cite{Burke:2013yra}.

In the past decade, sophisticated studies and tremendous progresses have been made in understanding the parton-medium interaction via various jet quenching observables in high-energy nucleus-nucleus collisions. These include nuclear modifications of single inclusive hadron production \cite{Bass:2008rv, Armesto:2009zi, Chen:2010te, Qin:2007rn}, di-hadron \cite{Zhang:2007ja, Renk:2008xq, Majumder:2004pt, Chen:2016vem} and photon-hadron correlations \cite{Renk:2006qg, Zhang:2009rn, Qin:2009bk, Li:2010ts}, full jet production and substructure \cite{Qin:2010mn, Dai:2012am, Wang:2013cia, Chang:2016gjp}, as well as heavy-flavor meson production \cite{Andronic:2015wma, Uphoff:2011ad, He:2011qa, Gossiaux:2010yx, Nahrgang:2013saa, Cao:2013ita, Cao:2015hia, Cao:2016gvr, Song:2015ykw, Das:2015ana}. However, challenges remain in achieving a fully consistent picture about the dynamical evolution and nuclear modification of hadron spectra of different flavors in different collision systems and energies.
One important phenomenon to understand is the flavor dependence of jet quenching. The comparable nuclear modification factors $R_\mathrm{AA}$ of $D$ mesons and light flavor hadrons at high $p_\mathrm{T}$ as measured at RHIC and LHC seem to contradict the conventional expectation from the color and mass dependences of parton energy loss \cite{Adare:2010de, Adamczyk:2014uip, ALICE:2012ab}.
The recent phenomenological observation \cite{Andres:2016iys} that the jet quenching parameter $\hat{q}/T^3$ extracted from experimental data on light hadron suppression depends only on the colliding energy but not on the centrality is also hard to understand if one considers temperature (or density) dependence of $\hat{q}$. 
%One challenge is the quantitative understanding and extraction of how jet-medium interaction strength depends on the properties of the traversed QGP %medium as well as the propagating jets. 

To resolve these difficulties and achieve a complete understanding of parton-medium interaction, its flavor, temperature and energy dependences, a comprehensive and consistent jet quenching framework is needed to simulate the evolution of both light and heavy partons inside the hot and dense QCD medium for various collision systems and energies, and to understand the available jet quenching data. Recently, we have developed a Linear Boltzmann Transport (LBT) model for simulating the hard parton evolution in QGP. In the LBT model, jet-medium interactions for light and heavy flavor partons are treated on the same footing for both elastic and inelastic processes. The collision kernels for elastic and inelastic collisions are calculated from perturbative QCD, and the space-time evolution of the bulk medium is provided by hydrodynamical simulations. One of the important features of the LBT model is that it naturally incorporates the jet energy, medium temperature and flow dependences of jet-medium interaction within the perturbative QCD approach. We report in this paper that our LBT model is able to simultaneously describe the nuclear modification factors of single inclusive charged hadrons and $D$ mesons for different centralities at both RHIC and the LHC energies. We also extract the jet quenching parameter $\hat{q}$, including its dependences on medium temperature and jet flavor.

\textit{A Linear Boltzmann Transport (LBT) model for parton evolution in medium} ---
We describe the scattering of energetic partons inside a thermal medium using the Boltzmann equation. Within the LBT model \cite{Wang:2013cia,Cao:2016gvr}, the phase space evolution of an incoming parton (denoted as ``$a$") can be described by
\begin{eqnarray}
  \label{eq:boltzmann1}
  p\cdot\partial f_a(x,p)=E (\mathcal{C}_\mathrm{el}+\mathcal{C}_\mathrm{inel}),
\end{eqnarray}
where $\mathcal{C}_\mathrm{el}$ and $\mathcal{C}_\mathrm{inel}$ denote collision integrals for elastic and inelastic scatterings, respectively.

For an elastic scattering process ($a+b\rightarrow c+d$),
\begin{eqnarray}
  \label{eq:Cel}
  \mathcal{C}_\mathrm{el}&=&\sum_{b,c,d}\frac{\gamma_b}{2E}\int \frac{d^3 p_b}{(2\pi)^3 2E_b}\int\frac{d^3 p_c}{(2\pi)^3 2E_c}\int\frac{d^3 p_d}{(2\pi)^3 2E_d}\nonumber\\
&\times&  \Big\{f_c(\vec{p}_c)f_d(\vec{p}_d)\left[1\pm f_a(\vec{p}) \right]\left[1\pm f_b(\vec{p}_b)\right]-\nonumber\\
&&f_a(\vec{p})f_b(\vec{p}_b)\left[1\pm f_c(\vec{p}_c) \right]\left[1\pm f_d(\vec{p}d)\right]\Big\} S_2(s,t,u)\nonumber\\
&\times& (2\pi)^4\delta^{(4)}(p+p_b-p_c-p_d)|\mathcal{M}_{ab\rightarrow cd}|^2,
\end{eqnarray}
where $\gamma_b$ represents the spin-color degeneracy of parton ``$b$" and the ``$\pm$" signs denote the Bose-enhancement and Pauli-blocking effects. In Eq. (\ref{eq:Cel}),
%\begin{eqnarray}
% \label{eq:S2}
$S_2(s,t,u)=\theta(s\ge2\mu_\mathrm{D}^2)\theta(-s+\mu_\mathrm{D}^2\le t\le -\mu_\mathrm{D}^2)$ 
%\end{eqnarray}
is introduced to regulate the collinear ($u,t\rightarrow 0$) divergence of the matrix element $|\mathcal{M}_{ab\rightarrow cd}|^2$ where $\mu_\mathrm{D}^2=g^2T^2(N_c+N_f/2)/3$ is the Debye screening mass. In our work, light partons are assumed to be massless, and for heavy flavor quarks we take $m_c=$1.27~GeV and $m_b=$4.19~GeV. The leading-order matrix elements are used for elastic scattering processes.
%Note that if ``1" is a heavy quark, we let ``3" to be the outgoing heavy quark and ``2" and ``4" to be the thermal and recoiled light partons; if ``1" is a light parton, the outgoing parton which carries the larger energy is denoted as ``3" and the other as ``4".
The above equation contains both gain and loss terms for $f_a$, from the latter we may obtain the elastic ``scattering rate" for parton ``$a$" as
\begin{eqnarray}
 \label{eq:rate2}
 \Gamma_\mathrm{el}^a &=&\sum_{b,c,d}\frac{\gamma_b}{2E}\int \frac{d^3 p_b}{(2\pi)^3 2E_b}\int\frac{d^3 p_c}{(2\pi)^3 2E_c}\int\frac{d^3 p_d}{(2\pi)^3 2E_d}\nonumber\\
&\times& f_b(\vec{p}_b)\left[1\pm f_c(\vec{p}_c) \right]\left[1\pm f_d(\vec{p}_d)\right] S_2(s,t,u)\nonumber\\
&\times& (2\pi)^4\delta^{(4)}(p+p_b-p_c-p_d)|\mathcal{M}_{ab\rightarrow cd}|^2.
\end{eqnarray}
With this rate, the probability of elastic scattering for parton ``$a$" in each small time step $\Delta t$ can be obtained: $P^a_\mathrm{el}=1-e^{-\Gamma^a_\mathrm{el}\Delta t}$. The jet transport coefficients $\hat{q}_a$ and  $\hat{e}_a$ (energy loss rate) due to elastic scattering may be obtained from the above integral weighted by the transverse momentum broadening or energy loss of parton ``$a$".

To include the inelastic process of medium-induced gluon radiation, we first calculate the average number of emitted gluons from a hard parton in each time step $\Delta t$:
\begin{equation}
 \label{eq:gluonnumber}
 \langle N_g^a \rangle(E,T,t,\Delta t) = \Delta t \int dxdk_\perp^2 \frac{dN_g^a}{dx dk_\perp^2 dt}.
\end{equation}
In our work, the gluon radiation spectrum is taken from the higher-twist energy loss formalism \cite{Guo:2000nz,Majumder:2009ge,Zhang:2003wk}:
\begin{eqnarray}
\label{eq:gluondistribution}
%\frac{dN_g}{dx dk_\perp^2 dt}=\frac{2\alpha_s C_A \hat{q} P(x)}{\pi k_\perp^4} \left(\frac{k_\perp^2}{k_\perp^2+x^2 m^2}\right)^4{\sin}^2\left(\frac{t-t_i}{2\tau_f}\right),\nonumber\\
\frac{dN_g^a}{dx dk_\perp^2 dt}=\frac{2\alpha_s C_A \hat{q_a} P_a(x)k_\perp^4}{\pi \left({k_\perp^2+x^2 m^2}\right)^4} \, {\sin}^2\left(\frac{t-t_i}{2\tau_f}\right),
\end{eqnarray}
where $x$ and $k_\perp$ are the fractional energy and transverse momentum of the emitted gluon with respect to its parent parton, $\alpha_s$ is the strong coupling constant, $P_a(x)$ is the splitting function, and $\hat{q}_a$ is the transport coefficient due to elastic scattering. Eq. (\ref{eq:gluondistribution}) also contains the mass dependence for heavy quark radiative energy loss. $t_i$ denotes an ``initial time" or the production time of the ``parent" parton from which the gluon is radiated, and $\tau_f={2Ex(1-x)}/{(k_\perp^2+x^2m^2)}$ is the formation time of the radiation.
We apply a lower energy cut-off for the emitted gluon $x_\mathrm{min}=\mu_D/E$ in our calculation to regulate the divergence as $x\rightarrow 0$. Multiple gluon radiation is allowed during each time step, where the number $n$ of radiated gluons is assumed to follow a Poisson distribution with the mean taken as $\langle N^a_g \rangle$. Thus, the probability for the inelastic scattering process is $P^a_\mathrm{inel}=1-e^{-\langle N^a_g \rangle}$. Note for $g\rightarrow gg$ process, $\langle N^g_g\rangle/2$ is taken as the mean for the Poisson distribution to avoid double counting.

Given the probabilities of elastic and inelastic scattering processes at each time step, the total scattering probability is $P^a_\mathrm{tot}=P^a_\mathrm{el}+P^a_\mathrm{inel}-P^a_\mathrm{el}P^a_\mathrm{inel}$, which may be divided into two regions: pure elastic scattering with probability $P^a_\mathrm{el}(1-P^a_\mathrm{inel})$ and inelastic scattering with probability $P^a_\mathrm{inel}$.
Using these probabilities and the information about the propagating jet parton and the medium profiles at each time step, a Monte-Carlo (MC) method is applied to determine whether a given jet parton is scattered with the medium constituents, and whether the scattering is elastic or inelastic.
If a scattering happens, the branching ratio $\Gamma^a_{\mathrm{el},i}/\Gamma^a_\mathrm{el}$ is utilized to determine the scattering channel, and the corresponding differential rate [in  Eq. (\ref{eq:rate2})] is used to sample the four-momenta of outgoing partons.
If the scattering is inelastic, we first sample the number $n$ of radiated gluons using Poisson distribution, and then obtain the momenta of these $n$ gluons according to their differential spectra [Eq. (\ref{eq:gluondistribution})].
We treat the inelastic radiative contribution as $2\rightarrow 2+n$ process and the energy-momentum conservation is respected accordingly in kinematics.

\begin{figure}[tb]
  \epsfig{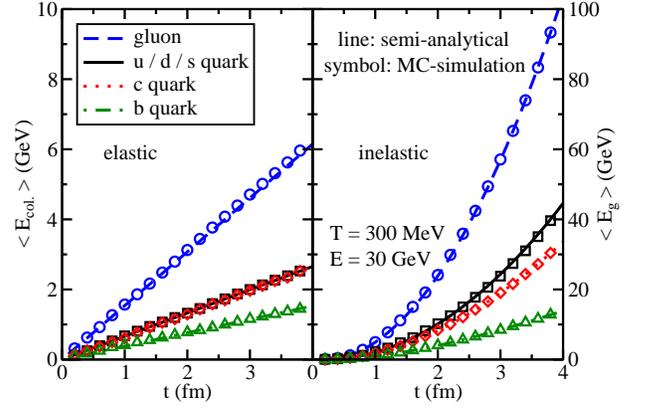}
  \caption{(Color online) Parton energy loss due to elastic and inelastic interactions in a static medium: semi-analytical calculation vs. Monte-Carlo simulation.}
 \label{fig:plot-eLoss}
\end{figure}

As a verification of our LBT model, the MC results of the cumulative energy loss for different parton species as a function of time
is compared to the semi-analytical calculation in Fig. \ref{fig:plot-eLoss}  for elastic scatterings (left panel) and inelastic processes (right panel)  (energy taken away by radiated gluons). Here we use a fixed strong coupling constant $\alpha_s=0.3$ and a static medium with temperature $T=300$~MeV. For a consistent comparison, we restore the energy of the leading parton back to the initial energy (30~GeV) after each evolution time step and also fix the initial time $t_i$ in Eq. (\ref{eq:gluondistribution}) to be $0$ in our LBT model since the semi-analytical calculation does not automatically include their variations during the time evolution. One can see that our MC simulations give rise to an elastic energy loss that increases linearly with time, and the slopes are in agreement with those from semi-analytical calculations of $\hat{e}$ (1.54~GeV/fm for gluon, 0.664~GeV/fm for light quark, 0.668~GeV/fm for charm quark and 0.382~GeV/fm for beauty quark). For inelastic processes, our LBT model shows a quadratic increase with time for the cumulative energy carried by the radiated gluons, and the MC results agree with the semi-analytical calculation as well.
A clear hierarchy in the energy loss for different parton species can also be observed. Gluons lose approximately 9/4 times as light quarks due to their different color charges. The energy loss of charm quarks is similar to that of light quarks for elastic scattering and is slightly smaller for inelastic process, since charm quark mass is small compared to the 30~GeV energy. However, the energy loss of beauty quarks is significantly smaller for both elastic and inelastic processes compared to light quarks and gluons.

\textit{Nuclear modifications of heavy and light flavor partons and hadrons} --- To study the evolution and modification of hard jet partons in heavy-ion collisions, we couple our LBT model to hydrodynamical simulation of the QGP medium. A (2+1)-dimensional viscous hydrodynamics model VISHNew \cite{Song:2007fn,Song:2007ux,Qiu:2011hf} with MC Glauber initial conditions is used in this work unless otherwise specified. The QGP formation time is set as $\tau_0=0.6$~fm and the shear-viscosity-to-entropy-density ratio ($\eta/s$=0.08) is tuned to describe the soft hadron spectra at both RHIC and the LHC. Jet partons are initialized as follows: the spatial profile is determined by binary collision distribution calculated from MC Glauber model, and the momentum distribution is calculated from leading-order perturbative QCD (LO pQCD), with the CTEQ parametrizations \cite{Lai:1999wy} for parton distribution functions and the EPS09 parametrizations \cite{Eskola:2009uj} for the nuclear shadowing effect. During the jet evolution, hydrodynamics provides the local temperature and flow velocity of the QGP fireball. To account for the medium flow effect, we first boost each jet parton into the local rest frame in which its energy and momentum are updated based on our LBT model, and then boost it back to the global collision frame. Jet partons are converted into hadrons on the freeze-out hypersurface of the fireball ($T_\mathrm{c}=165$~MeV). The hadronization of heavy quarks follows a hybrid model of fragmentation plus coalescence as in our earlier work \cite{Cao:2016gvr}, while the fragmentation of high $p_\mathrm{T}$ light partons to hadrons is simulated with \textsc{Pythia} MC model.

One crucial quantity controlling jet-medium interaction is the strong coupling constant $\alpha_s$, which enters into our calculation in a few places. Considering the fact that the variation of the medium temperature is small compared to that of the jet energy, we use a fixed coupling ($\alpha_s = 0.15$ obtained from the comparison to $R_\mathrm{AA}$ data, see below) for the interaction vertices connecting directly to thermal partons. But for the vertices connecting to hard partons, we take the running coupling as:
$\alpha_s(Q^2)={4\pi}/{[(11-2n_f/3)\mathrm{ln}(Q^2/\Lambda^2)]}$, 
with $Q^2=2ET$ and $\Lambda=0.2$~GeV. 
With such setup, our calculation naturally incorporates both energy and temperature dependences for jet-medium interaction. 

\begin{figure}[tb]
  \epsfig{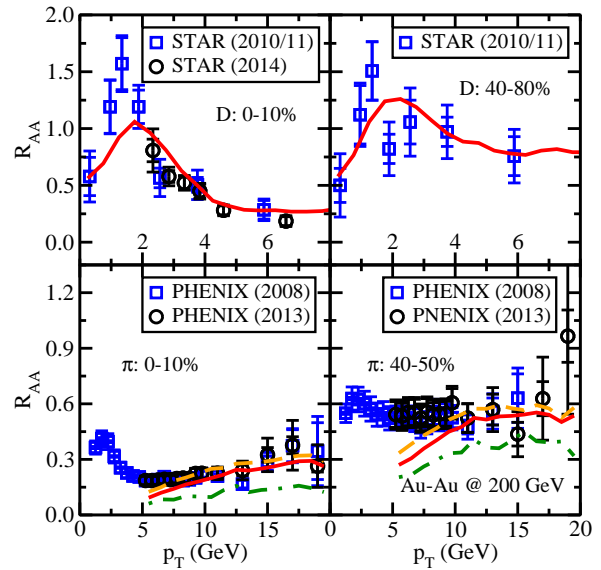}
  \caption{(Color online) Nuclear suppression factor $R_\mathrm{AA}$ of $D$ and $\pi$ spectra in 200~AGeV Au-Au collisions at RHIC. Data are taken from Refs. \cite{Adamczyk:2014uip,Xie:2016iwq,Adare:2012wg}. Contributions to $\pi$ spectra from quarks (upper curves with dashed lines), gluons (lower curves with dot-dashed lines) and the sum of them (middle curves with solid lines) are shown separately in the lower panels.}
 \label{fig:plot-RAA-both-200}
\end{figure}

\begin{figure}[tb]
  \epsfig{file=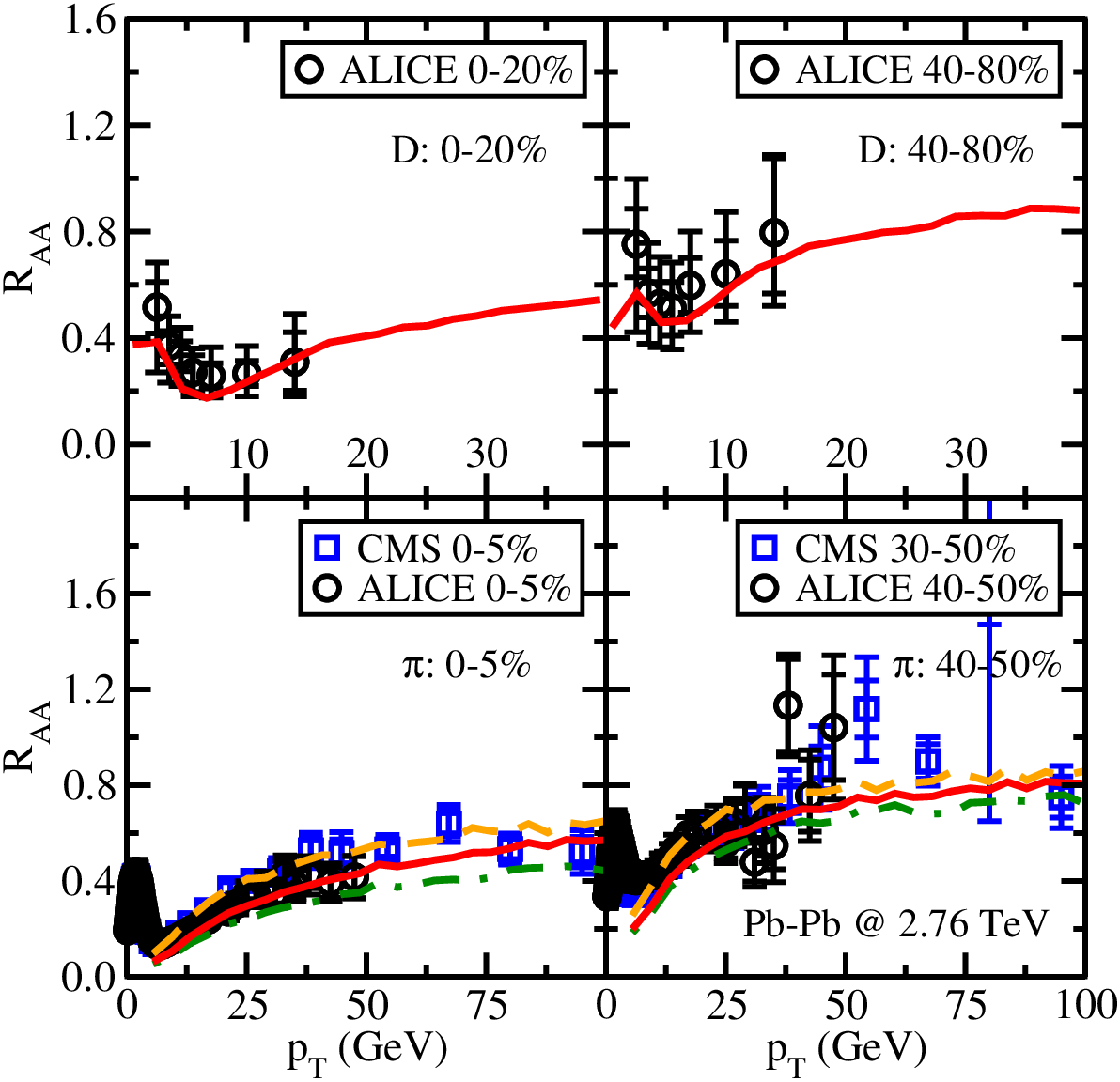, width=0.43\textwidth, clip=}
  \caption{(Color online) Nuclear suppression factor $R_\mathrm{AA}$ of $D$ and $\pi$ spectra in 2.76~ATeV Pb-Pb collisions at the LHC. Data are taken from Refs. \cite{ALICE:2012ab,CMS:2012aa,Abelev:2012hxa}. Notations of different curves are the same as in  Fig.~\ref{fig:plot-RAA-both-200}.}
 \label{fig:plot-RAA-both-2760}
\end{figure}

\begin{figure}[tb]
  \epsfig{file=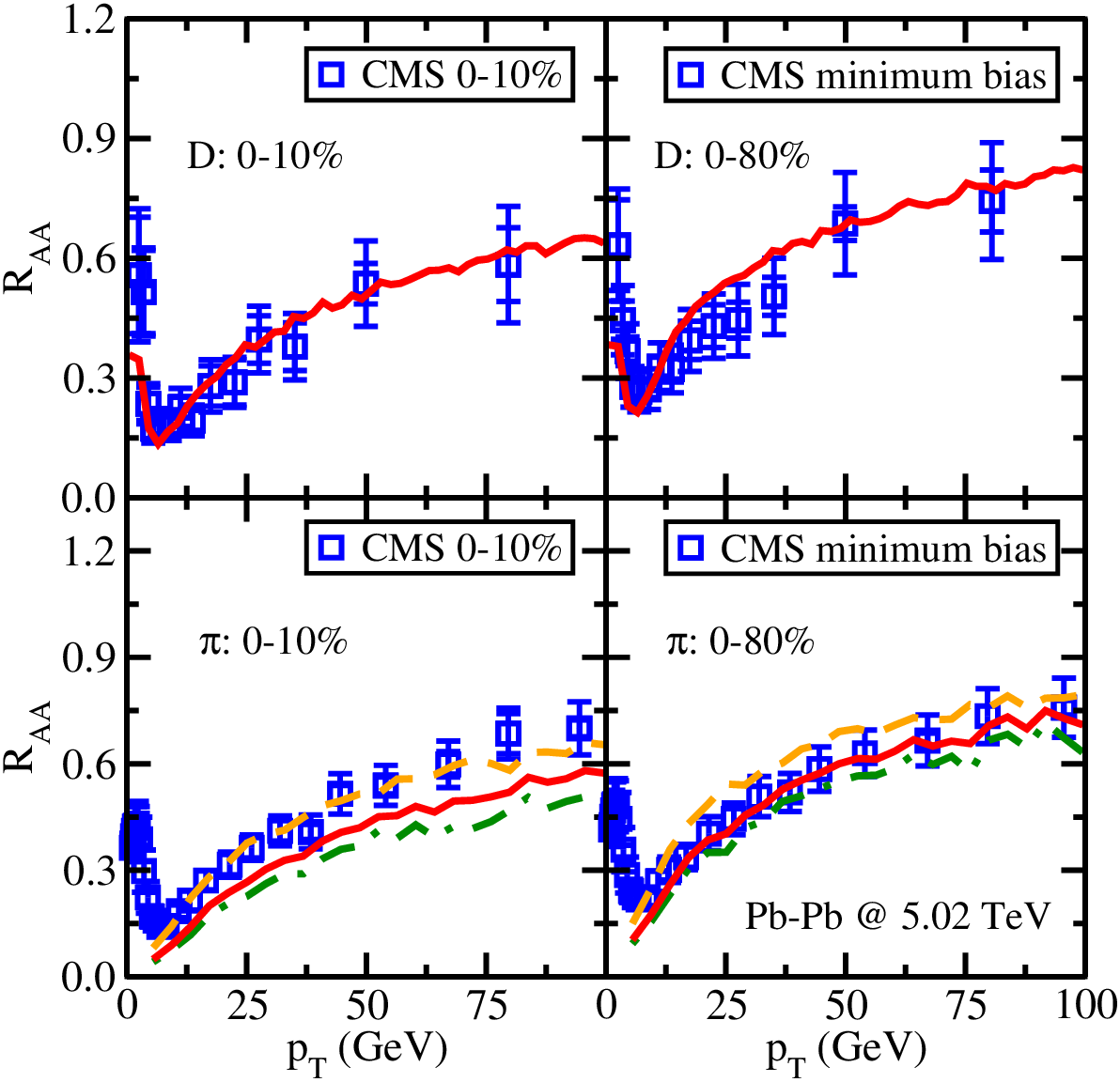, width=0.43\textwidth, clip=}
  \caption{(Color online) Nuclear suppression factor $R_\mathrm{AA}$ of $D$ and $\pi$ spectra in 5.02~ATeV Pb-Pb collisions at the LHC. Data are taken from Refs. \cite{CMS:2016nrh,Khachatryan:2016odn}.  Notations of different curves are the same as in  Fig.~\ref{fig:plot-RAA-both-200}.}
 \label{fig:plot-RAA-both-5020}
\end{figure}

We now present in Figs. \ref{fig:plot-RAA-both-200}, \ref{fig:plot-RAA-both-2760} and \ref{fig:plot-RAA-both-5020} the nuclear modification factors $R_\mathrm{AA}$ for $D$ and $\pi$ mesons in 200~AGeV Au-Au collisions at RHIC, 2.76~ATeV and 5.02~ATeV Pb-Pb collisions at the LHC. In all plots for $\pi$'s, the three curves denote $R_\mathrm{AA}$ for $\pi$ produced from quarks only (upper), gluons only (lower) and the sum of them (middle). Note that for 5.02~ATeV Pb-Pb collisions, the spacetime evolution of QGP profiles is provided by a (3+1)-dimensional viscous hydrodynamics model CLVisc \cite{Pang:2012he,Pang:2014ipa} in which $\tau_0=0.6$~fm, $\eta/s=0.08$ and $T_\mathrm{c}=165$~MeV are consistently employed. We also note that since our transport coefficients are obtained from LO pQCD calculation, we apply a $K$ factor for jet transport coefficients at low momentum in order to describe the low $p_\mathrm{T}$ heavy meson data. Here we use $K_p=1+A_p e^{-|\vec{p}|^2/2\sigma_p^2}$ with $A_p=5$ and $\sigma_p=5$~GeV following our previous study \cite{Cao:2016gvr}. One can see that with the inclusion of the energy and temperature dependences for jet-medium interaction in our LBT model, we obtain good descriptions of the nuclear modification factors for both heavy and light flavor hadrons in different centralities at different colliding energies.

\begin{figure}[tb]
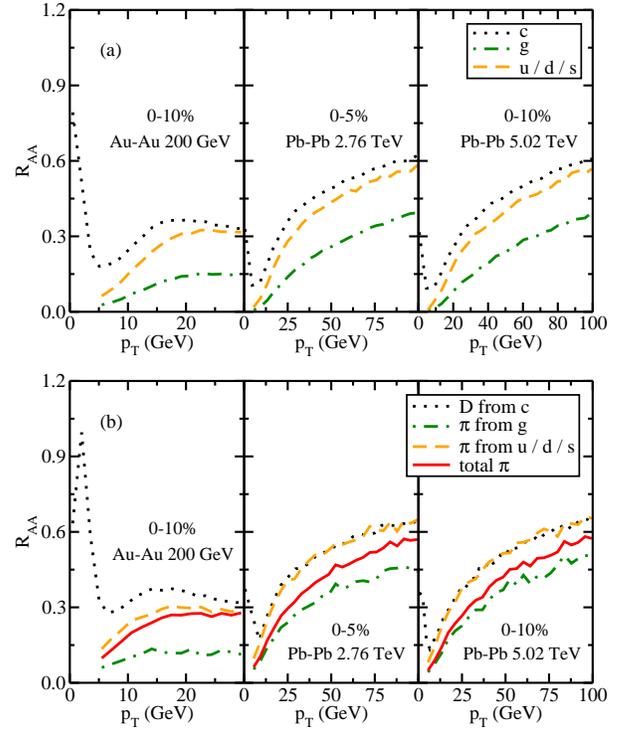

 \subfigure{\label{fig:plot-RAA-HLcompareP}
  \epsfig{file=plot-RAA-HLcompareP.eps, width=0.44\textwidth, clip=}}
 \subfigure{\label{fig:plot-RAA-HLcompareH}
  \epsfig{file=plot-RAA-HLcompareH.eps, width=0.44\textwidth, clip=}}
  \caption{(Color online) Flavor dependence of nuclear suppression factor $R_\mathrm{AA}$ from RHIC to the LHC energies for (a) parton and (b) hadron spectra.}
  \label{fig:plot-RAA-HLcompare}
\end{figure}

To better understand the flavor dependence of jet quenching, we compare in Fig. \ref{fig:plot-RAA-HLcompare} the nuclear modification of light and heavy flavor partons [Fig. \ref{fig:plot-RAA-HLcompareP}] and hadrons [Fig. \ref{fig:plot-RAA-HLcompareH}] for central RHIC 200~AGeV Au-Au collisions (left) and the LHC 2.76~ATeV (middle) and 5.02~ATeV (right) Pb-Pb collisions. As expected, $R^c_\mathrm{AA}>R^q_\mathrm{AA}>R^g_\mathrm{AA}$ is observed at the parton level due to their energy loss hierarchy as shown in Fig. \ref{fig:plot-eLoss}. However, we see that $R_\mathrm{AA}$ of $D$ mesons is very similar to that of $\pi$'s produced from light quark jets at the LHC. This is mainly caused by their hard initial spectra at the LHC energies that result in their quick rise of $R_\mathrm{AA}$ at high $p_\mathrm{T}$ together with the harder fragmentation function of charm quark than that of light quark. Furthermore, we observe the increasing gluon contribution to light hadron production from RHIC to the LHC energies which also leads to the different splittings of $R_\mathrm{AA}$ 
 for $D$ and $\pi$ mesons in these systems.

\begin{figure}[tb]
  \epsfig{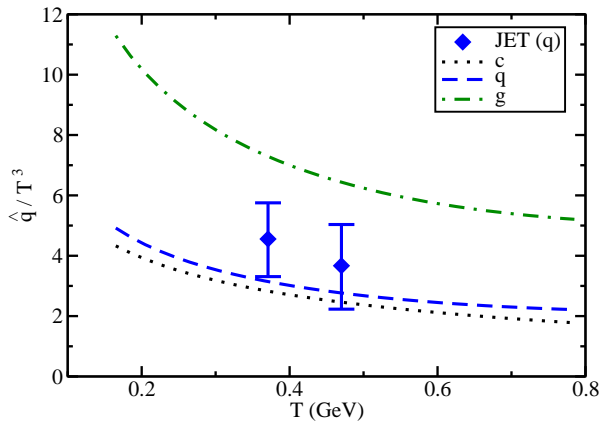}
  \caption{(Color online) Flavor and temperature dependences of $\hat{q}/T^3$ at $p=10$~GeV from our LBT model compared to that from JET (for light quark).}
  \label{fig:plot-qhat-T}
\end{figure}

One can extract the jet transport coefficient $\hat{q}$ from our model to data comparison for charm quark, light quark and gluon as functions of the medium temperature as shown in in Fig. \ref{fig:plot-qhat-T}. A decrease of $\hat{q}/T^3$ with temperature increasing from $T_\mathrm{c}$ is observed. Note that our result is consistent with the values for light quark at $p=10$~GeV obtained by the JET Collaboration \cite{Burke:2013yra}.

\textit{Summary} ---
We have extended the Linear Boltzmann Transport (LBT) model to include heavy and light flavor hard partons on the same footing when they undergo both elastic and inelastic interactions in the QGP medium. With a hydrodynamic description of the bulk medium evolution as the dynamic background, we have achieved good descriptions of the experimental data on nuclear modification of both light hadrons and $D$ mesons for different centralities at RHIC and the LHC energies. Using our LBT model constrained by the experimental data, we have quantitatively extracted the jet quenching parameter $\hat{q}/T^3$ with its dependences on jet flavor and medium temperature. In addition to the flavor dependence of parton energy loss, we have also found the importance of both initial parton spectra and fragmentation functions to the thorough understanding of the difference between the nuclear modifications of light and heavy flavor hadrons.

\textit{Acknowledgments} ---
We thank Long-Gang Pang for providing the hydrodynamics profiles for 5.02~ATeV Pb-Pb collisions and the computational resources from the Open Science Grid (OSG) and Duke University. We thank Yayun He and Wei Chen for helpful discussions. This work is supported in part by the Natural Science Foundation of China (NSFC) under Grant No. 11521064 and No. 11375072, by the Chinese Ministry of Science and Technology under Grant No. 2014DFG02050 and by the Major State Basic Research Development Program in China (No. 2014CB845404), by the Director, Office of Energy Research, Office of High Energy and Nuclear Physics, Division of Nuclear Physics, of the U.S. Department of Energy under Contract No. DE-AC02-05CH11231 and DE-SC0013460.

\bibliographystyle{h-physrev5}
\bibliography{SCrefs}

\end{document}